\documentclass[slac_one]{revtex4}
\usepackage{graphicx}
\usepackage{fancyhdr}
\newcommand{\mean}[1]{\langle#1\rangle}
\newcommand{\GeV}{{\,\rm GeV}}
\newcommand{\ee}{\mbox{$e^+e^-$~}}
\newcommand{\diff}{\mathrm{d}}
\newcommand{\alp}{\alpha_s}
\newcommand{\alps}{\alpha_s}
\def\MSbar{$\overline{\mbox{\rm MS}}$}

\begin{document}


\title{Review of Power Corrections in DIS}

\author{T. Kluge}

\affiliation{DESY, Notkestr. 85, D-22607 Hamburg, Germany }

\begin{abstract}
An overview is given of analyses in DIS at HERA which confront the predictions of power corrections with measured data.
These include mean values and distributions of 2-jet as well as 3-jet event shape variables and jet rates. 

\end{abstract}

\maketitle
\thispagestyle{fancy}

\section{Introduction}
Event shape variables are known to exhibit rather large hadronisation effects, not treated by 
purely perturbative calculations.
It has been shown \cite{Webber:1994zd} that the size of these corrections varies as 
$(\Lambda/Q)$, with $Q$ being the hard scale of the process, for most of the event shape observables.
This article focuses on power corrections in the approach initiated by Dokshitzer and Webber
\cite{Dokshitzer:1995zt}, since these are the most complete theory and often used in comparisons to data.
There is a wealth of studies performed in \ee annihilation which
gives support to this concept, cf.\ to contribution {\bf R002} in these proceedings.   
It is interesting to extend these studies to deep-inelastic scattering (DIS),
in order to investigate the universality of the ansatz and to check for possible 
modifications of the hadronisation process due to the presence of a proton remnant.
After power corrections to mean values of event shapes variables became available \cite{Webber:1995ka},
the H1 and later also the ZEUS Collaboration published analyses which corroborated this concept \cite{Adloff:1997gq,Chekanov:2002xk}.
A comprehensive review on the subject of event shapes can be found in \cite{Dasgupta:2003iq}. 

When comparing the $ep$ scattering to \ee annihilation, one moves from an s-channel to a t-channel exchange,
with the negative four momentum transferred $Q$ corresponding to the center of mass energy $\sqrt s$.
At HERA a large range of the scale is available in a single experiment, typically stretching $ 5 < Q< 115\GeV$.
The theoretical treatment is complicated by
initial state singularities related to the incoming proton, which are absorbed in the parton density functions (pdf).
In order to reject the proton dissociation part of an event (the remnant), the event shapes are calculated
in the Breit frame of reference, where the separation between particles from the hard scattering and
the remnant is clearest.
The boost to the Breit frame is determined by the event kinematics and the boosted particles are separated into
hemispheres with pseudorapidity $\eta<0$ (current hemisphere) and $\eta>0$ (remnant hemisphere).
Usually only particles of the current hemisphere enter the event shape definition, where this 
hemisphere resembles to some extend one half of an \ee event.
In order to ensure infrared safety at all orders a minimal energy in the current hemisphere is applied
as part of the observable's definition, e.g.\ $E_{CH}>Q/10$.

\section{Event Shape Variables}
The observables covered in the following may be separated into three categories: 2-jet event shapes, 3-jet event shapes and jet rates.
The term ``$n$-jet'' denotes that at least $n$ particles in the final state are needed for a non vanishing value of the variable.
Note that particles from the proton remnant is not included in $n$.  
Common examples for 2-jet event shapes are thrust, jet broadening, jet mass and the C-parameter.
For this class of observables the most advanced theory predictions are available, therefore many analyses concentrate
on those.
3-jet variables which have been proposed are the out-of-plane momentum and the azimuthal correlation.
While the theoretical predictions are yet less complete for these variables, additional insights to power corrections
are possible due to sensitivity to hadronisation from a gluon and applications to hadron-hadron collisions.

Closely related to event shapes are jet rates which make use of a jet clustering algorithm, such as the $k_t$ or the JADE algorithm.
Here the algorithm is applied on particles on both hemispheres and the $n$-jet rate is defined as the maximum value of
the cut off parameter $y_{\rm cut}$ where the event is clustered to $n+1$ jets ($+1$ denotes the proton remnant).

A particular distinction made for 2-jet event shapes is only possible for DIS: between ones making use of the exchanged boson axis
 (mostly from a virtual photon) in their definition and those without that reference axis, as for the definitions used in  
\ee annihilation.
Namely for thrust and jet broadening both variants are investigated, whereby the explicit use of the boson direction
 implies sensitivity to radiation into the remnant hemisphere through recoil effects on the current 
quark~\cite{Dasgupta:2002dc}.

The thrust variable $\tau$\/ with respect to the boson axis is defined as
\begin{eqnarray}
\tau=1-T\quad\mathrm{with}\quad  T&=&\frac{\sum_h|\vec p_{z,h}|}{\sum_h|\vec p_h|} \ ,
  \label{eqn:thrust} 
\end{eqnarray}
and the thrust variable $\tau_C$\
\begin{eqnarray}
\tau_C=1-T_C\quad\mathrm{with}\quad T_C&=&\max_{\vec n_T} \frac{\sum_h|\vec p_h\cdot \vec n_T|}{\sum_h|\vec p_h|} \ ,
  \label{eqn:thrustc}
\end{eqnarray}
 where the direction $\vec n_T$
 maximises the sum of the longitudinal momenta of all particles
in the current hemisphere along this axis.

The Jet Broadening is defined as 
\begin{eqnarray}
  B&=&\frac{\sum_h|\vec p_{t,h}|}{2\sum_h|\vec p_h|} \ .
  \label{eqn:bparameter}
\end{eqnarray}

The squared Jet Mass is calculated as
\begin{eqnarray}
  \rho&=&\frac{(\sum_h E_h)^2 - (\sum_h\vec p_h)^2}{(2\sum_h|\vec p_h|)^2}.
  \label{eqn:jetmass} 
\end{eqnarray}
In the following the symbol $\rho_0$ is used, which indicates that
in the above definition the hadrons are treated as massless, replacing
the energy $E_h$ by the modulus of the 3-momentum $|\vec p_h|$.
This adjustment is made since the theoretical predictions assume the partons to be massless.
Mass effects can be huge, especially for the jet mass observable.

The $C$-Parameter
\begin{eqnarray}
  C&=&\frac{3}{2}\frac{\sum_{h,h'}|\vec p_h||\vec p_{h'}|\sin^2\theta_{hh'}}
                      {(\sum_h|\vec p_h|)^2} \ ,
  \label{eqn:cparameter}
\end{eqnarray}
where $\theta_{hh'}$ is the angle between particles $h$ and $h'$.
Note that in all the definitions the momenta are defined in the Breit frame and the sums extend
over all particles in the current hemisphere.

The out-of-event plane momentum is defined as
\begin{eqnarray}
K_{\textrm{out}}={\sum_h}' |p_h^{\textrm{out}}|.
\end{eqnarray}
Results are presented in terms of $K_{\textrm{out}}/Q$.
 Here $p_h^{\textrm{out}}$ is the out-of-plane momentum of the hadron $h$ with the event plane defined
 to be formed by the proton
momentum $\vec P$ in the Breit frame and the unit vector $\vec n$ which enters the definition of thrust major:
 \begin{eqnarray}
T_M=\max_{\vec n}\frac{1}{Q}{\sum_h}' |\vec p_h\cdot \vec n|,\qquad \vec n\cdot \vec P=0
\end{eqnarray}
To avoid measurements in the beam region, the sum indicated by ${\sum_h}'$ extends over all hadrons with pseudo-rapidity 
 $\eta<3$ in the Breit frame.
 The restriction to only the current hemisphere ($\eta<0$), as for the 2-jet shapes, would be too restrictive because of the
 extended phase space available for three partons. For reasons
discussed in \cite{Banfi:2001ci}, only events with $p_t\sim Q$ should be selected, which is accomplished by a cut on the (2+1)-jet
resolution $y_2$ defined by the $k_t$ clustering algorithm: $0.1<y_2<2.5$.

The azimuthal correlation between the hadrons labelled $h$ and $i$ is defined as
\begin{displaymath}
\chi=\pi-|\phi_h-\phi_i| \ ,
\end{displaymath}
where the observable is constructed by summing over all hadron pairs with a weight
\begin{displaymath}
w=\frac{p_{th}p_{ti}}{Q^2}.
\end{displaymath}
The azimuth in the Breit frame of hadron $h$ is denoted by $\phi_h$.

Predictions of mean values of 2-jet event shapes are available up to next-to-leading order (NLO) in the strong coupling,
 together with power corrections.
Mean values of 2-jet rates for the JADE and the $k_t$ algorithm have been compared to NLO calculations,
 however the power corrections for this observables are not completely known.
Soft gluon resummations to next-to-leading log precision (NLL) have been performed for distributions of 2-jet and 3-jet event shapes.
These have been matched to NLO fixed order predictions for the 2-jet event shapes only, power corrections are known for both.
Distributions of jet rates can only be compared to NLO alone, as no matched soft gluon resummation nor power corrections are available yet.

\section{Mean Values}
In the Dokshitzer and Webber approach, 
the mean value of an event shape variable is modified through non-perturbative
effects by an additive constant~\cite{Dokshitzer:1995zt} 
\begin{equation}
  \mean{F} = { \mean{F}^{\textrm{\scriptsize pQCD}}}+a_F{ \mathcal{P}},
\end{equation}
where $a_F$ is of order one and can be calculated perturbatively. 
The power correction term $\mathcal{P}$ is assumed to be universal for all
event shape variables.
It is proportional to $1/Q$ and 
evaluated to be
\begin{equation}
  \mathcal{P}=\frac{16}{3\pi}\mathcal{M}\frac{\mu_I}{Q}
  \left [\alpha_0(\mu_I)-\alpha_s(Q)-\frac{\beta_0}{2\pi}
    \left(\ln \frac{Q}{\mu_I}+\frac{K}{\beta_0}+1\right)
    \alpha_s^2(Q) \right ] \; ,
  \label{pformula}
\end{equation}
where $\beta_0=11-2\,n_f/3$, 
$K = 67/6 -\pi^2/2-5\, n_f/9$, and
$n_f=5$ is the number of active flavours. 
The so-called Milan factor $\mathcal{M}\simeq0.95$ in the \MSbar\ scheme
ensures the universality at the two-loop level~\cite{Dasgupta:1998xt}.
This ansatz results in only one single non-perturbative parameter 
$\alpha_0 = \mu_I^{-1}\int_0^{\mu_I} \alpha_\mathrm{eff}(k)\,\mathrm{d}k$,
being the first moment of the effective coupling integrated over the low scale region
up to the matching scale $\mu_I$.
$\alpha_0$ has to be determined from data and is expected to be around $0.5$.  

For the perturbative part of the calculation the NLO programs DISENT \cite{Catani:1997vz} and DISASTER++ \cite{Graudenz:1997gv} have been
used.
The uncertainty due to missing higher orders of the perturbative series is by convention 
estimated by a variation of the renormalisation scale by a factor of two ($\mu_r=Q/2$ and $\mu_r=2Q$). 

\subsection{2-jet Event Shapes}

\begin{figure}
\includegraphics[width=100mm]{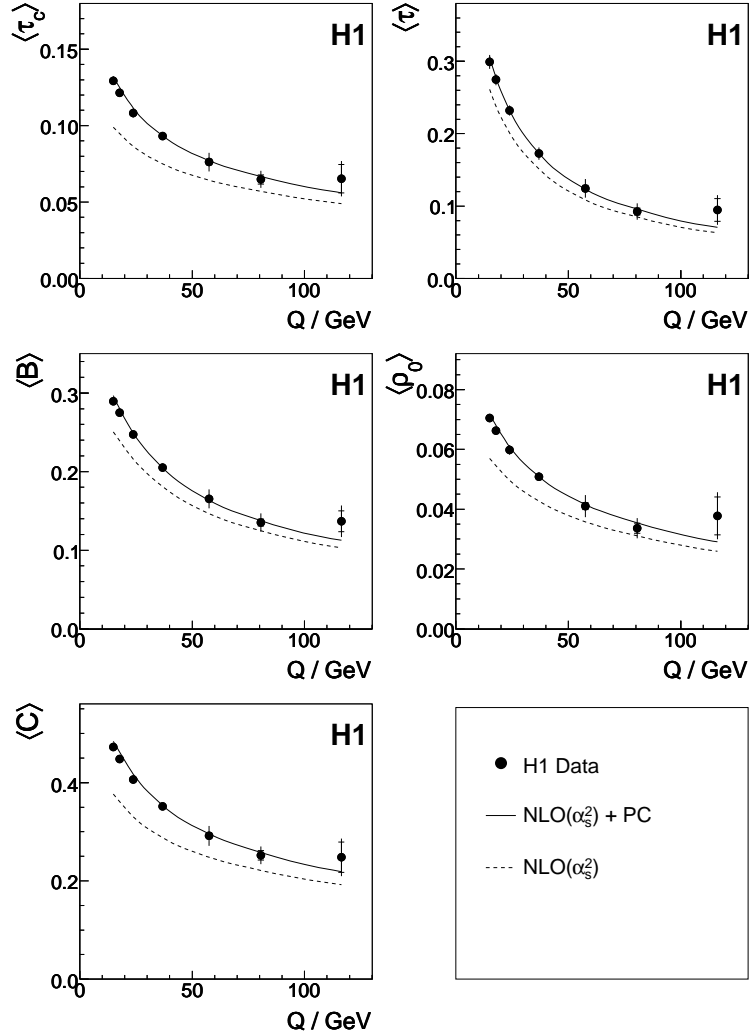}
\caption{Mean values of event shape variables corrected to the hadron level as a function of the scale
    $Q$.
    The data, presented with statistical errors (inner bars)
    and total errors (outer bars),
    are compared with  the results of NLO QCD fits including power corrections (PC).
 The dashed curves show the NLO QCD contribution to the fits \cite{Aktas:2005tz}.}
\label{f5}
\end{figure}
The H1 Collaboration measured the mean values of five event shapes as function of $Q$  \cite{Aktas:2005tz}, shown in Fig.~\ref{f5}.
A steady decrease of the means with rising $Q$ is observed and a
good description of the data is obtained by a fit of DISASTER++  supplemented by power corrections \cite{Dasgupta:1998ex,Dasgupta:1998xt}.
For comparison the fixed order contribution alone is also given, which demonstrates the importance of the power corrections,
 especially for the non-photon-axis variables $\tau_c$, $\rho_0$ and C-Parameter. 
Earlier predictions of the jet broadening \cite{Dokshitzer:1998pt} were not able to describe the data because of a more involved perturbative non-perturbative interplay of this observable. 

\begin{figure}
\includegraphics[width=100mm]{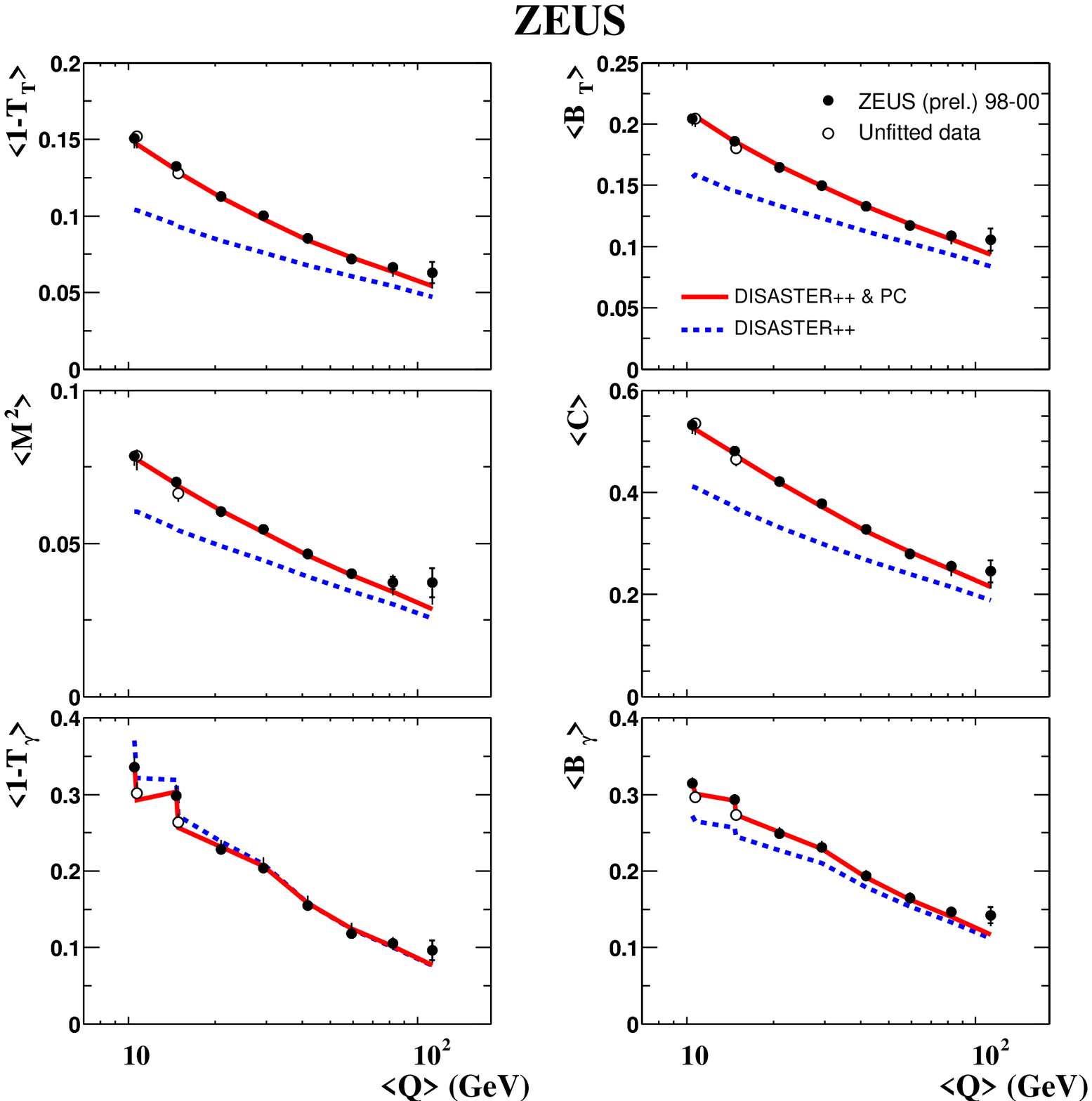}
\caption{The mean values of event-shape variables as a function of $Q$ . The solid lines are the results of the fit to the data of the predictions
of the sum of NLO pQCD calculations from DISASTER++ and the power corrections. The dashed lines are the DISASTER++ contribution to the fit alone \cite{Everett:2004fg}.}
\label{f6}
\end{figure}
\begin{figure}
\includegraphics[width=110mm]{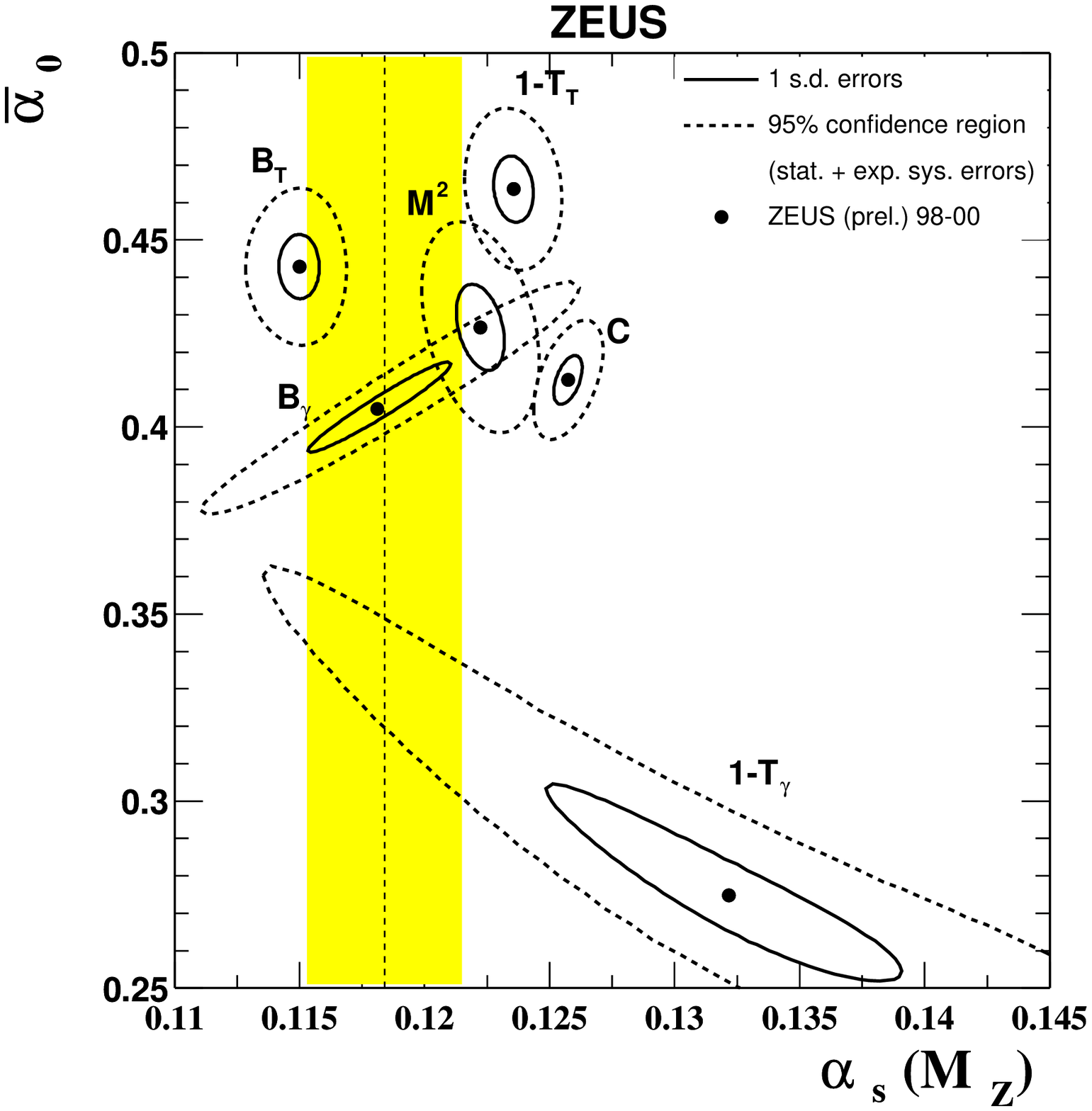}
\caption{Extracted parameter values for ($\alpha_s$,$\alpha_0$) from fits to the mean values of the shape variables.
The vertical line and the shaded area indicate the world average of $\alpha_s(m_Z)$ \cite{Everett:2004fg}.}
\label{f7}
\end{figure}
\begin{figure}
\includegraphics[width=100mm]{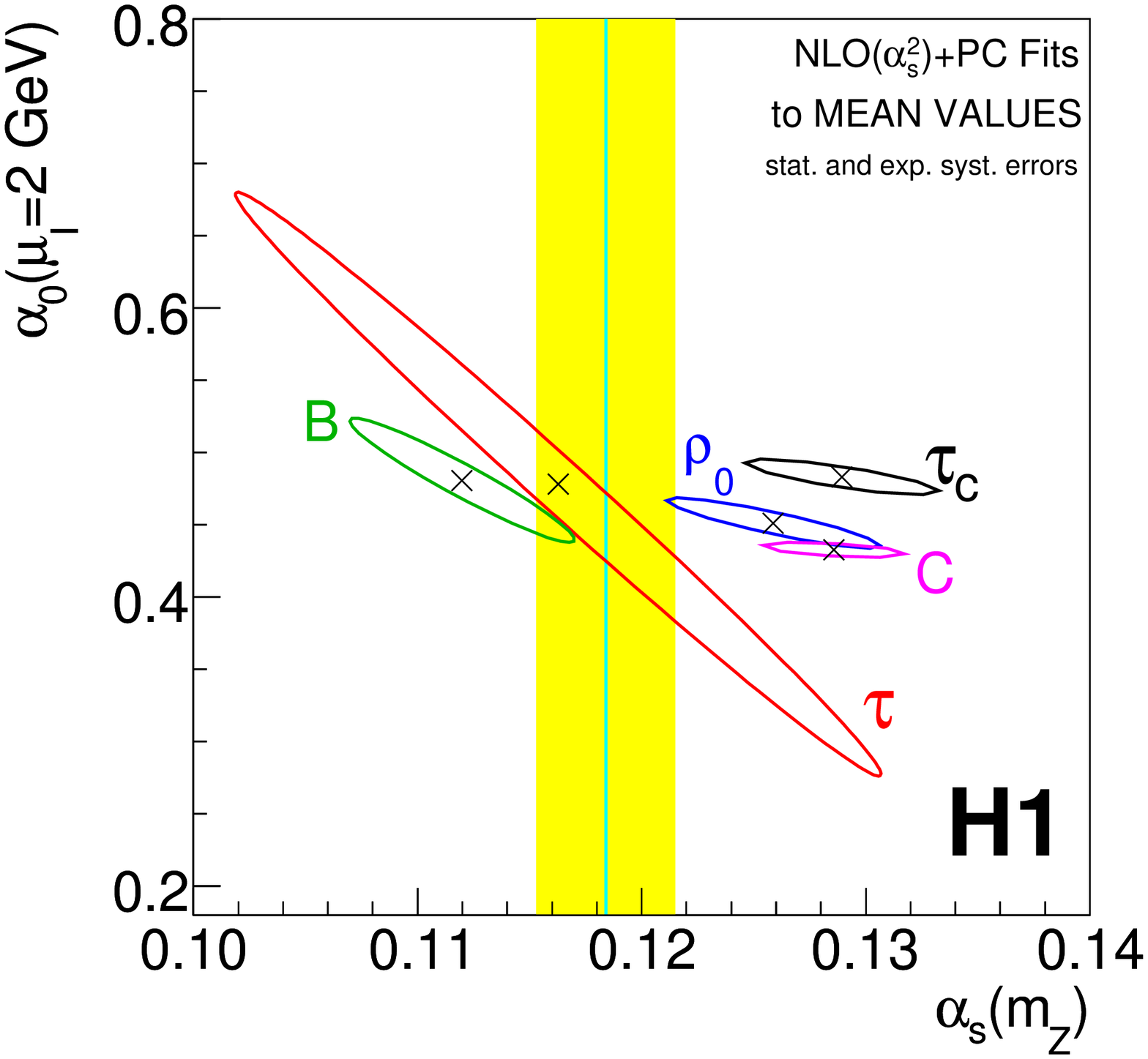}
\caption{Results of fits to the mean values of
    $\tau$, $B$, $\rho_0$, $\tau_C$ and the $C$-parameter 
    in the $(\alpha_s,\alpha_0)$ plane.
    The $ 1\sigma$ contours
    correspond to $\chi^2 = \chi^2_{\rm min}+1$,
    including statistical and experimental systematic uncertainties \cite{Aktas:2005tz}.
    The value of $\alpha_s$ (vertical line) and its uncertainty (shaded band) are taken from \cite{Bethke:2004uy}.}
\label{f8}
\end{figure}

Similar results are obtained by the ZEUS Collaboration \cite{Everett:2004fg}, shown in Fig.~\ref{f6}.
Note the logarithmic abscissa and the differing labels for the event shapes used compared to H1 where
$1-T_T$ corresponds to $\tau_c$, $1-T_\gamma$ to $\tau$, $B_\gamma$ to $B$ and $M^2$ to $\rho_0$.
Surprisingly the fitted power correction for the thrust with respect to the boson axis ($1-T_\gamma$)
 turns out to be slightly negative. 
The fitted
values of $\alpha_s(m_Z)$ and $\alpha_0$ are shown in Fig.~\ref{f7} and \ref{f8}.
The 1-sigma contours denote experimental errors alone, which are only at the few percent level.
These errors are mainly caused by imperfect knowledge of the electromagnetic energy scale of the calorimeter, which enters
the boost to the Breit frame of reference.
Not shown are the sizeable theoretical errors (dominated by scale uncertainties due to missing higher orders) of
about ten percent.
In both analyses $\alpha_0$ is found to be universal at the $10\%$ level,
and the spread of the fitted $\alpha_s$ happens to be larger than expected from the experimental errors.
There are some differences observed with the results from both collaborations, e.g. the values of $\alpha_0$
fitted by H1 cluster around a slightly  higher value than those from ZEUS.
However, the overall pattern looks quite similar, with $\tau$ having a large error and a huge correlation between
the fitted parameters.
Also the observables which make not use of the virtual boson axis ($\tau_c$,$\rho_0$ and C-parameter)
prefer rather high values of $\alpha_s(m_Z)$ (compared to the world mean), in contrast to the broadening $B$,
which exhibits the lowest values in both analyses.

\begin{figure}
\includegraphics[width=140mm]{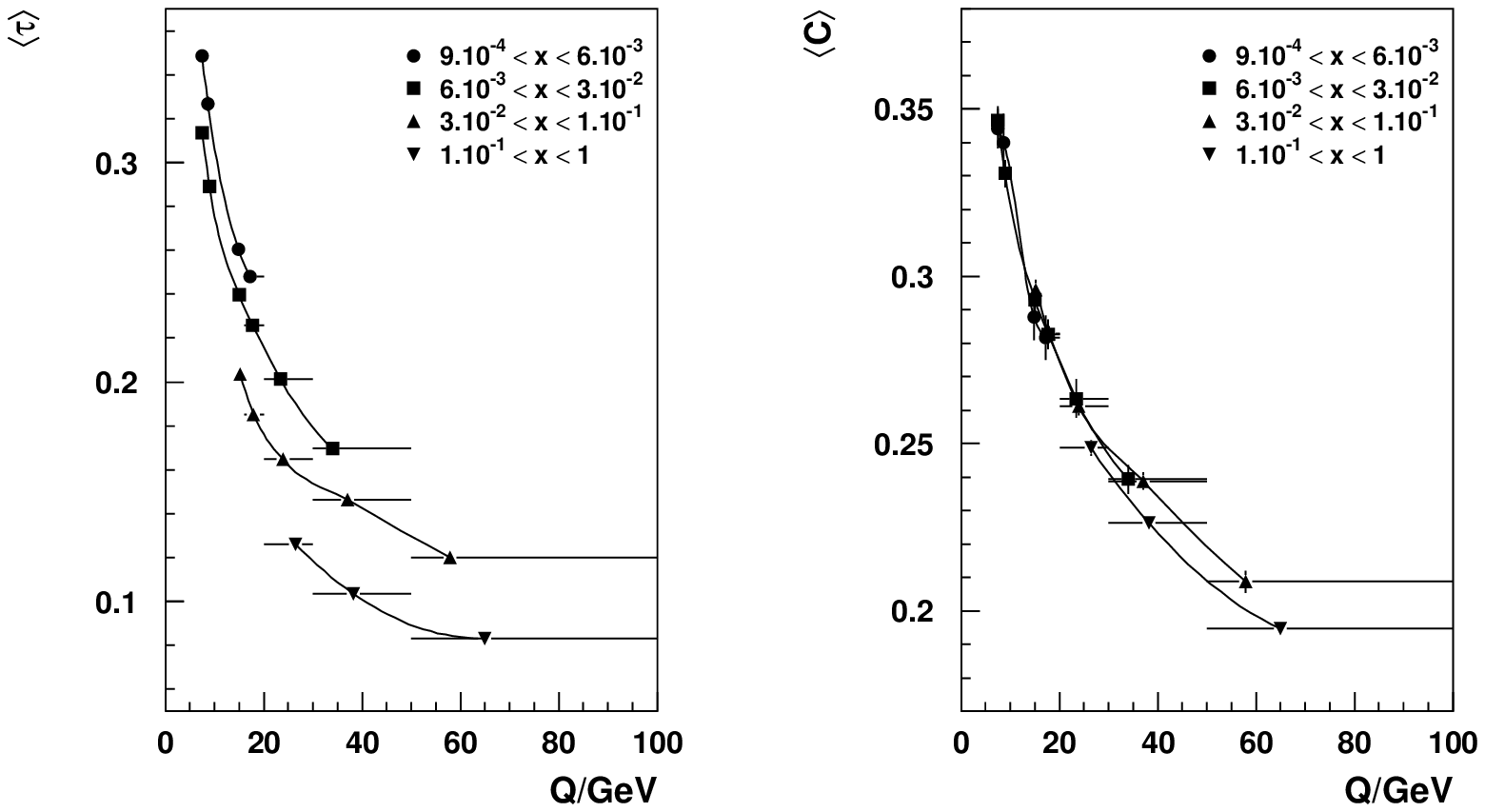}
\caption{Mean values of $\tau$ (left) and $C$ (right) versus $Q$ in four different bins of $x$ calculated with DISENT. The
lines connect the means belonging to the same $x$ bin \cite{Adloff:1999gn}.}
\label{f9}
\end{figure}

In contrast to \ee annihilation, where the perturbative coefficients are constant, in DIS they depend on $x$ according to
the parton density functions and the accessible $x$-range at different values of $Q$.
Variables defined with respect to the boson axis, like $\tau$ are expected to show a stronger dependence on $x$,
 which is demonstrated in Fig.~\ref{f9}.
While the perturbative contributions are expected to depend on $x$, the power corrections must not,  
 in order of the concept to work. 
A former analysis by the ZEUS Collaboration \cite{Chekanov:2002xk} observed an $x$ dependent power correction parameter $\alpha_0$ for $\tau$,
$\tau_c$ and even $\rho_0$,
however this observation was not confirmed by more recent investigations.

\subsection{Jet Rates}
\begin{figure}
\includegraphics[width=130mm]{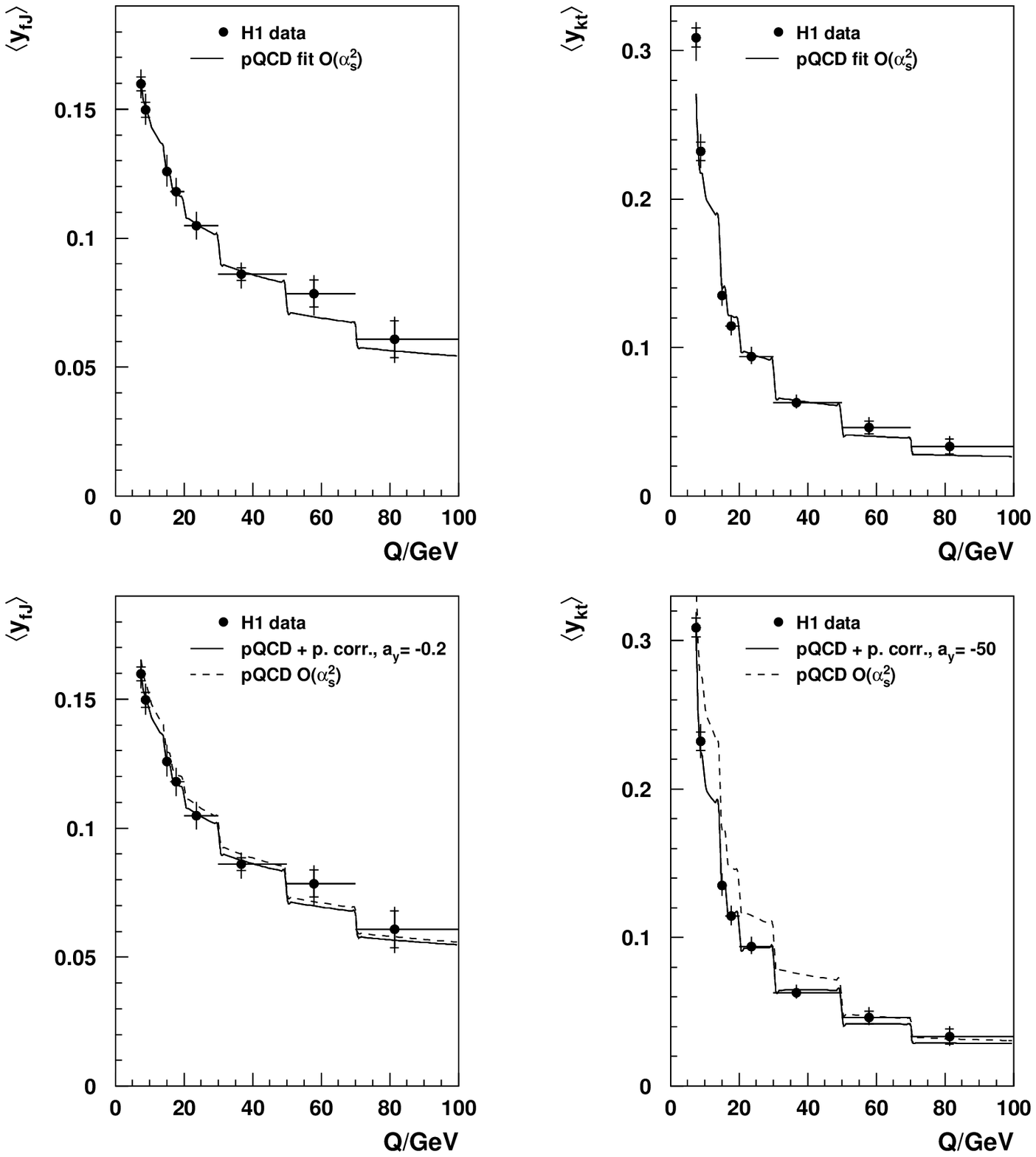}
\caption{Mean values of $y_{fJ}$ and $y_{kt}$ as a function of $Q$ The error bars represent statistical and systematical uncertainties.
Upper part: The full line corresponds to a fit of the pQCD calculation without power corrections. Lower part: The full line corresponds to
a power correction fit according to the Dokshitzer-Webber approach. 
The dashed line shows the pQCD contribution of DISENT in these fits \cite{Adloff:1999gn}.}
\label{f10}
\end{figure}

Jet rates show in general smaller hadronisation corrections than event shapes.
Power corrections $\propto (\Lambda/Q)^{2p}$ have been proposed for the JADE and the $k_t$ jet algorithms, 
with suggested values of $p=1/2$ and $p=1$, respectively.
Reliable calculations of the coefficients $a_f$ have not been presented yet.
In an H1 analysis  \cite{Adloff:1999gn} an attempt was made to fit this coefficients together with $\alpha_s$ and $\alpha_0$, 
the resulting theory curves are shown in Fig.~\ref{f10}. 
While data are described quite well by the fit,
the hadronisation corrections turn out to be negative and the
fitted values of $\alpha_s$ are unphysical.
Since the effect of hadronisation is so small for this class of variables, even more precise
experimental data could help to resolve this issue. 

\section{Distributions}
Compared to mean values, the study of the spectra of event shape variables offers several advantages.
Firstly, the shape of the distributions is governed by QCD and thus offers more information available for fits.
Secondly, when fitting it is possible to restrict the range used to an interval where the theory
prediction is reliable.
\subsection{2-jet Event Shapes}
Towards low values of the shape variables there are terms due to soft gluon radiation, which become dominant, but are
not included in the NLO prediction used for the mean values.
A significant improvement of the description is obtained with a resummation of these soft gluons, matched to the NLO part.
Such a calculation is available in DIS for the 2-jet variables $\tau$, $B$, $\tau_c$, $\rho_0$ and C-Parameter.
\begin{figure}
\includegraphics[width=64mm]{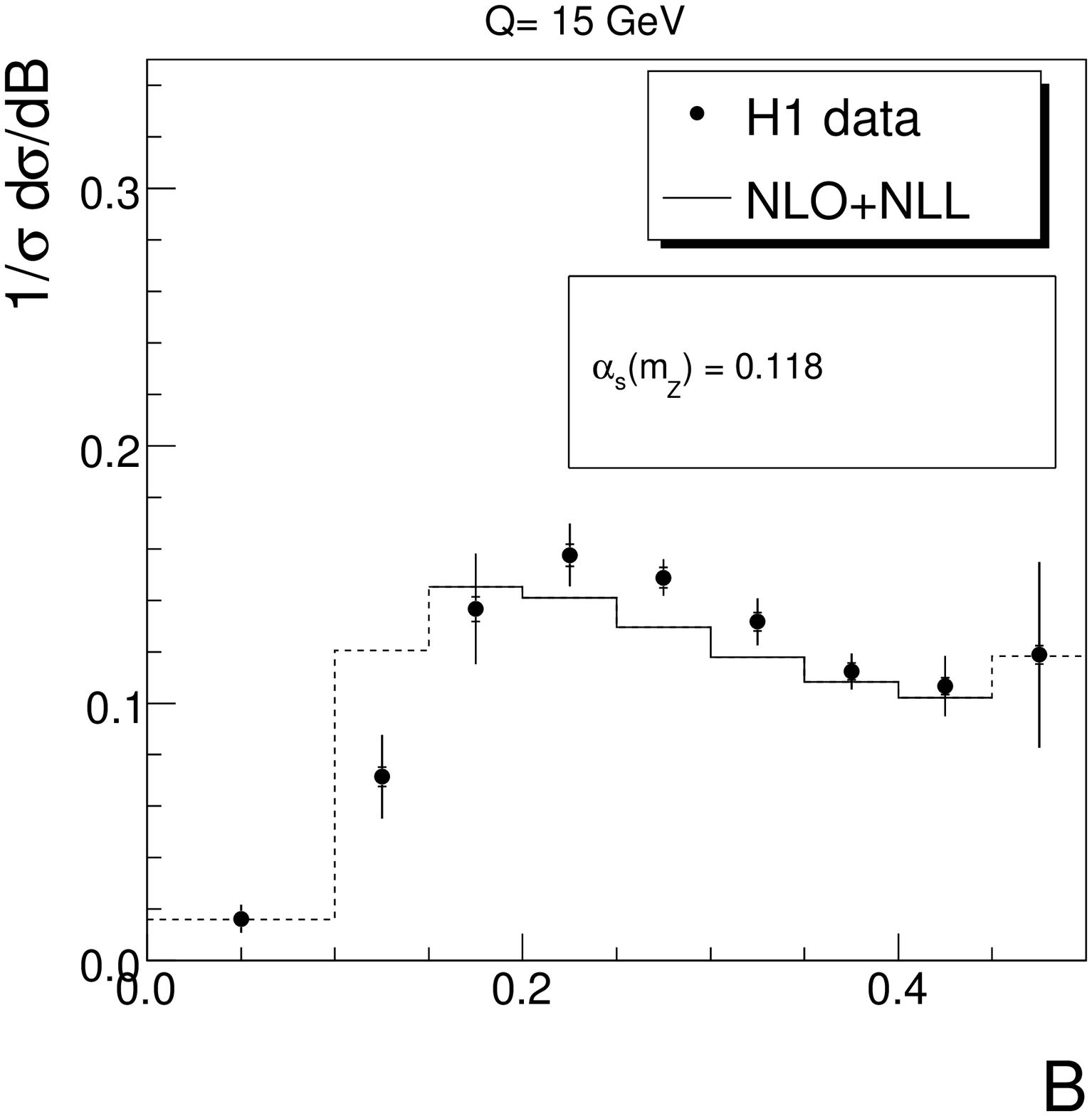}
\includegraphics[width=64mm]{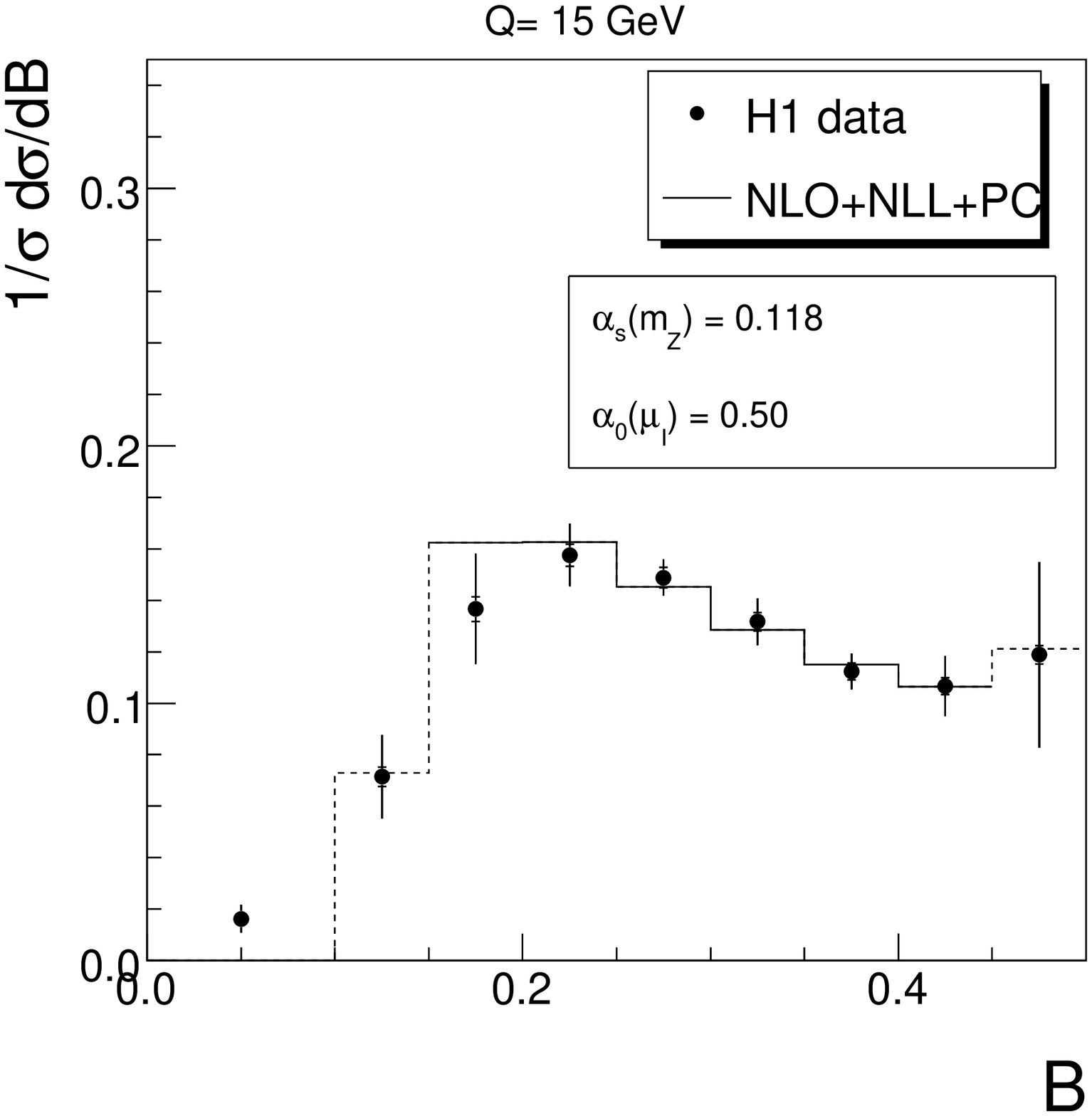}
\caption{Spectra of the Jet Broadening for $\mean{Q}=15\GeV$. The prediction by NLOJET++ and soft gluon resummation (left) needs to be 
supplemented by power corrections in order to describe the data (right).}
\label{f11}
\end{figure}
Fig.~\ref{f11} shows exemplarily that a large part of the central region of a spectrum is nicely
reproduced by the theory, even at a rather low scale of $Q=15\GeV$, where hadronisation effects are 
not small.

The hadronisation correction in the Dokshitzer and Webber approach results in a shift for the differential distributions
\begin{equation}
\frac{1}{\sigma_{\mathrm{tot}}}\frac{\diff\sigma(F)}{\diff F}=\frac{1}{\sigma_{\mathrm{tot}}}\frac{{ \diff\sigma^{\mathrm{pQCD}}}(F-a_F\mathcal{ P})}{\diff F},
\end{equation} 
with the same coefficient $a_F$ and power correction $\mathcal{P}$ as for the mean values.
This shift cannot be valid over the whole spectrum,
at low values of $F$ it may be applied only for $F \gg a_F\mathcal{ P} \sim
\mu_I/Q$~\cite{Dokshitzer:1997ew}.
Moreover, at large values of $F$  higher order corrections are substantial and the perturbative
calculation is not reliable.
In consequence for each analysis one has to decide on
 the interval used for fits, depending on the observable and $Q$.
For the upper bounds the values given in \cite{Dasgupta:2002dc} are used,
 while for the lower bounds of the fit intervals different methods are applied.
In case of the jet broadening the picture is more complicated, as the shift is supplemented by a squeeze \cite{Dasgupta:2001eq}.

\begin{figure}
\includegraphics[width=130mm]{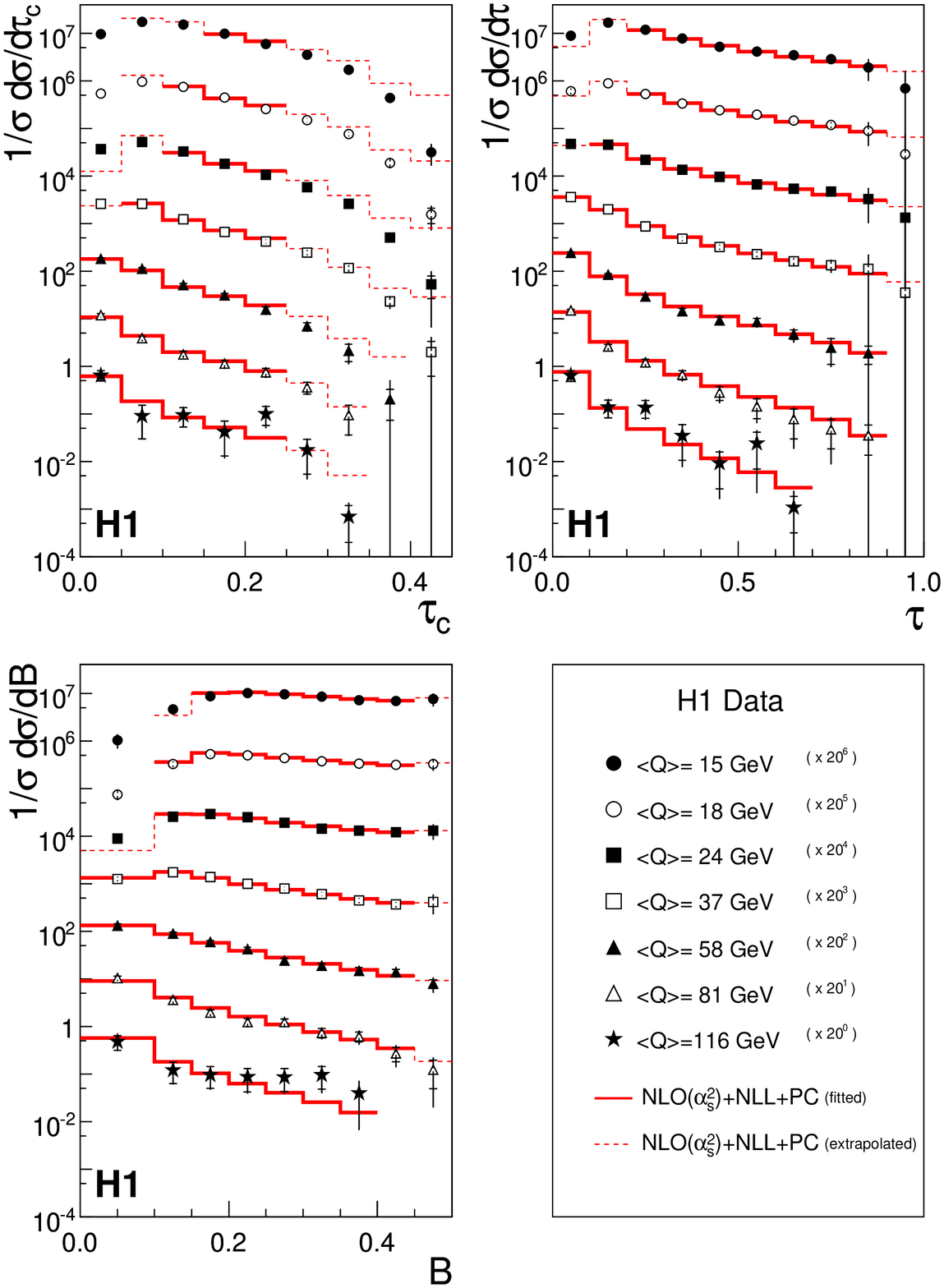}
\caption{ Normalised event shape distributions corrected to the hadron level
    for $\tau_C$, $\tau$ and $B$.
    The data are presented with statistical errors (inner bars)
    and total errors (outer bars).
    The measurements are compared with fits based on a NLO QCD calculation including 
    resummation (NLL) and supplemented by power corrections (PC).
    The fit results are shown as solid lines and are extended as
    dashed lines to those data points which are not included in the QCD fit  \cite{Aktas:2005tz}.}
\label{f1}
\end{figure}

\begin{figure}
\includegraphics[width=130mm]{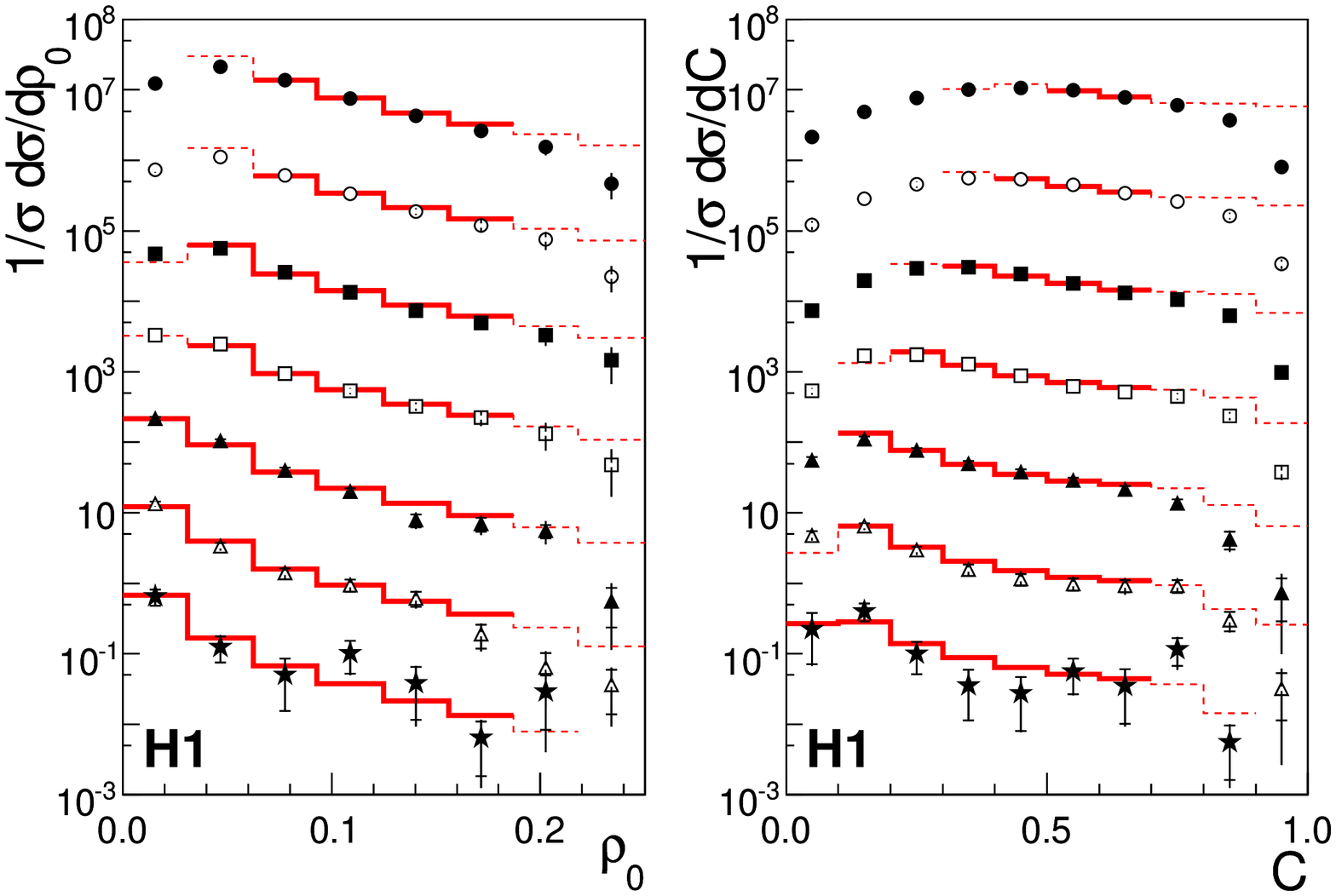}
\caption{Normalised event shape distributions corrected to the hadron level
    for $\rho_0$ and the $C$-parameter. 
   For details see the caption of Fig.~\ref{f1}.}
\label{f2}
\end{figure}

\begin{figure}
\includegraphics[width=110mm]{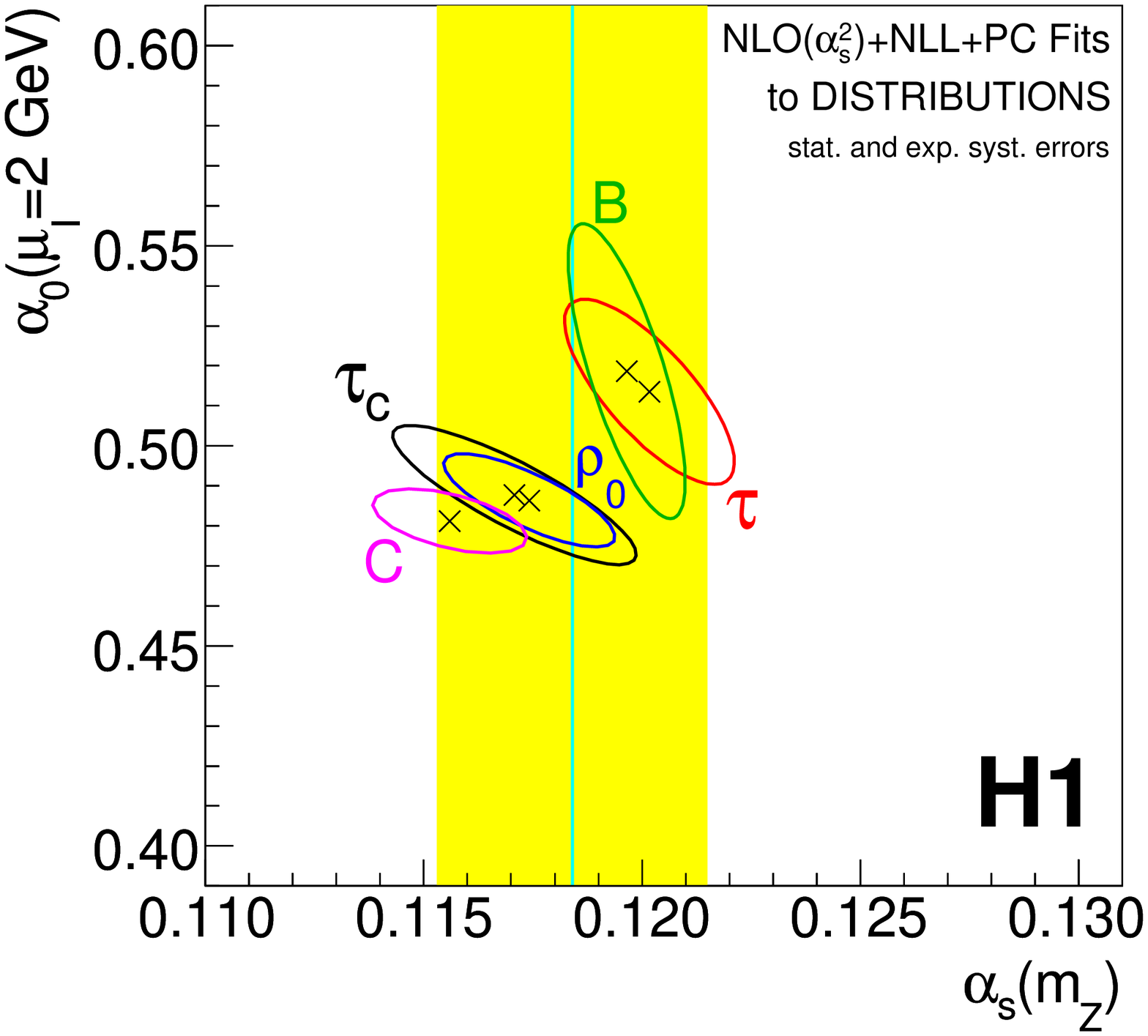}
\caption{Fit results to the differential distributions of
    $\tau$, $B$, $\rho_0$, $\tau_C$ and the $C$-parameter 
    in the $(\alpha_s,\alpha_0)$ plane \cite{Aktas:2005tz}.
    The $ 1\sigma$ contours
    correspond to $\chi^2 = \chi^2_{\rm min}+1$,
    including statistical and experimental systematic uncertainties.
    The value of $\alpha_s$ (vertical line) and its uncertainty (shaded band) are taken from \cite{Bethke:2004uy}.}
\label{f3}
\end{figure}

The calculations shown in the following were performed using DISASTER++ for the fixed order part and the
DISRESUM package \cite{Dasgupta:2002dc} containing the resummed contributions and the power corrections.
In a recent publication  \cite{Aktas:2005tz} the H1 Collaboration has measured the differential distributions for the five event shapes quoted above, shown together
with a QCD fit in Fig.~\ref{f1} and \ref{f2}
The data cover a wide range of $\langle Q \rangle = 15 - 116~\GeV$. 
Except for the highest $Q$ bins, 
the precision of the measurements is not statistically limited.
For each variable the shape of the spectra changes considerably with increasing
$Q$, becoming narrower and evolving towards low values.
The results of the combined fit for $\alpha_0$ and $\alp(m_Z)$ are displayed in 
Fig.~\ref{f3}.
The quality of the fits, expressed in terms of $\chi^2$ per degree of freedom,
is found to be between 0.5 and 1.4.
It was checked that the results are stable against the omission of data points at the edges
of the fit intervals.
For all event shape variables consistent values for $\alp(m_Z)$ and $\alpha_0$ are found,
with a maximum difference of about two standard deviations between $\tau$ and $C$.
The fitted results for the five observables appear to be grouped for those with ($\tau$, $B$) and without
($\tau_c$, $\rho_0$ and $C$) referring to the boson axis in their definition.
 A strong negative correlation between  $\alp(m_Z)$ and  $\alpha_0$ is observed for all
variables. 
The values of the strong coupling $\alpha_s(m_Z)$ are in good agreement with
the world average~\cite{Bethke:2004uy}, shown for comparison as the
shaded band.
The non-perturbative parameter $\alpha_0\simeq 0.5$ is confirmed to be
universal within $10\%$. 
A combined analysis of all event shape variables yields
\begin{eqnarray}
  \alps(m_Z) & = &
      0.1198 \pm 0.0013\ ({\rm exp})\ ^{+0.0056} _{-0.0043}\ ({\rm theo}) \ ,
      \nonumber \\[1ex]
  \alpha_0   & = &
      0.476 \pm  0.008 \ ({\rm exp})\ ^{+0.018} _{-0.059}\  ({\rm theo}) \ ,
      \nonumber
\end{eqnarray}
where the theoretical error is derived from the renormalisation scale uncertainty.
The total errors are dominated by this renormalisation scale uncertainty, which suggests
that missing higher order terms in the perturbative calculation are important.

\begin{figure}
\includegraphics[width=110mm]{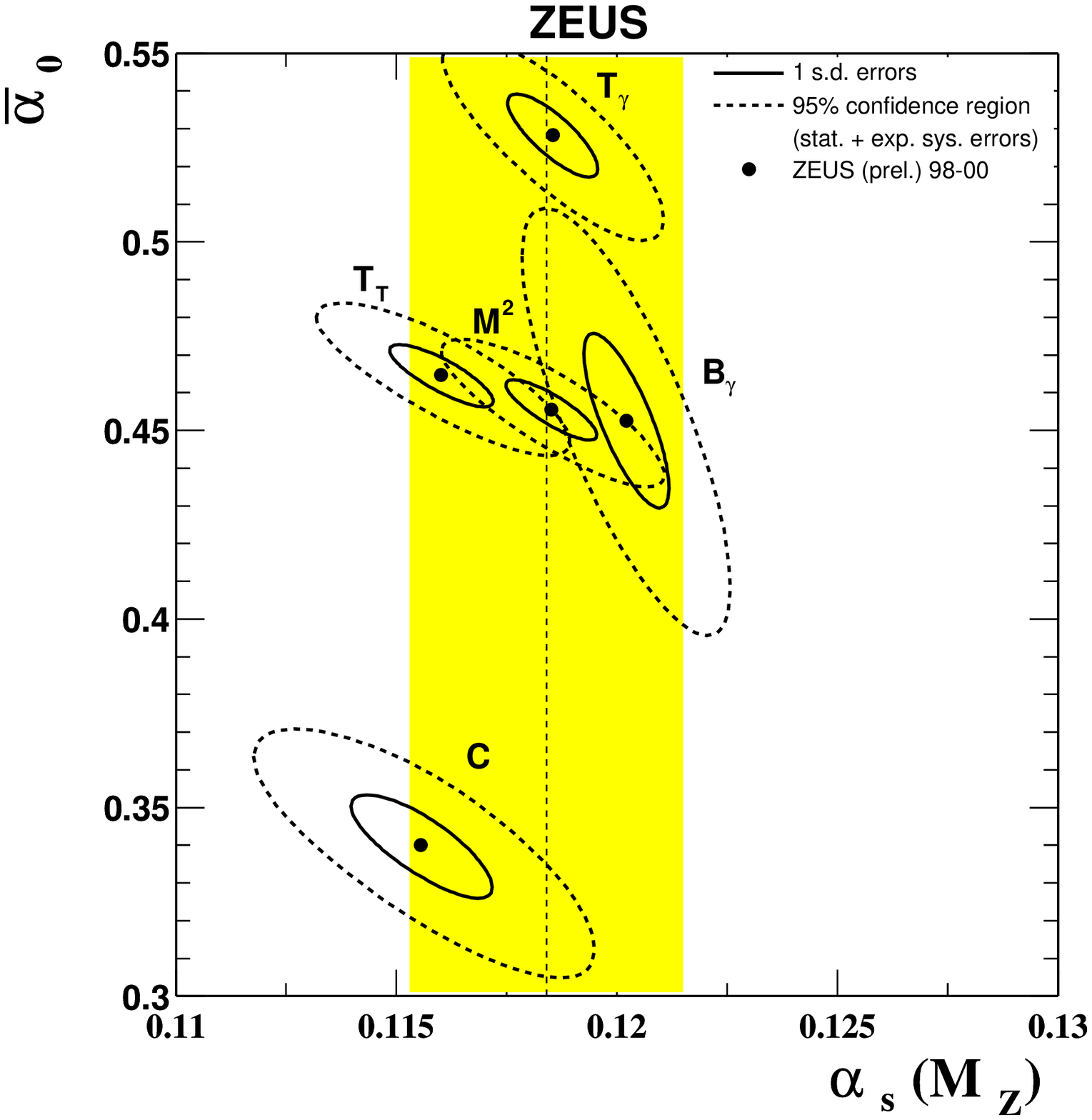}
\caption{Extracted parameter values for ($\alpha_s$,$\alpha_0$) from fits to differential distributions of shape variables \cite{Everett:2004fg}.
The vertical line and the shaded area indicate the world average of $\alpha_s(m_Z)$.}
\label{f12}
\end{figure}

A less consistent picture with respect to the universality of $\alpha_0$ is provided by 
preliminary results from the ZEUS Collaboration, shown in Fig.~\ref{f12}
While the fitted values of $\alp(m_Z)$ are found to agree with the world average, a rather
low value of $\alpha_0$ is obtained for the C-Parameter, not covered by the experimental errors.
However, only for $\tau$ and $B$ the value of $\chi^2$ per degree of freedom is around unity,
 for the remaining variables it is around five.
So the question arises whether the boundaries of the fit were stretched beyond the region where
the theory is predictive. 
It is surprising that almost no (or even negative) hadronisation was found for the thrust (wrt. the boson axis)
in case of the ZEUS mean value analysis ($\alpha_0<0.3$ in Fig.~\ref{f6}), while the differential distributions
 extracted from the same data to not show this property (with $\alpha_0>0.5$). 

The event shape spectra used for the fit appear to be consistent with those from H1,
but there are some differences in the treatment of the systematic errors.
Moreover, there is some freedom when choosing options in the matching of the fixed order and 
the resummed part of the calculation and in deciding on proper fit intervals.
Still, both analyses agree with respect to several findings: 
$\alp(m_Z)$ is compatible with the world mean (which is not the case for the event shape mean values),
$\alp(m_Z)$ and $\alpha_0$ have a negative correlation coefficient and 
the best description (quantified by $\chi^2$/d.o.f.) is obtained for the boson axis variables.
In order to draw  more decisive conclusions about the validity of power corrections a consistent handling
 of the theory parameters, fit intervals and systematic errors would be desirable.
Also, reduced theory uncertainties would help a lot,
e.g.\ by a refined combination procedure \cite{Jones:2003yv} or even a NNLO calculation. 

\begin{figure}
\includegraphics[width=90mm]{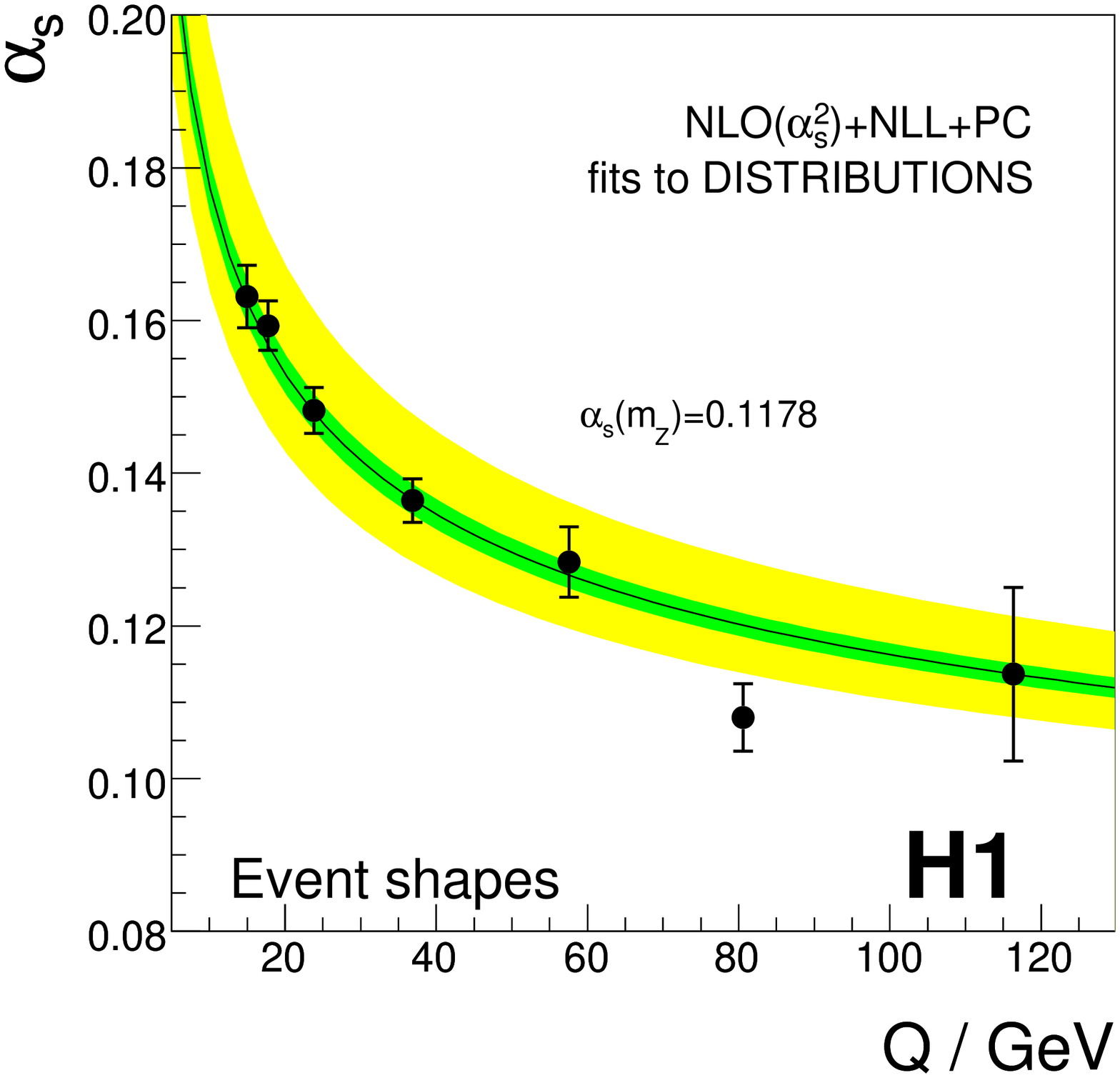}
\caption{The strong coupling $\alps$ as a function of the scale $Q$.
The individual fit results, shown
as points with error bars, are obtained
from fits to the differential distributions in $\tau_C$, $\tau$, $B$, $\rho_0$ and $C$
within each $Q$ bin.
The errors represent the total experimental uncertainties.
For each event shape observable a value of 
$\alps(m_Z)$ is indicated in the plot, determined
from a fit to the $\alps(Q)$ results using the QCD renormalisation
group equation.
The corresponding fit curves are shown as full lines.
The shaded bands represent the uncertainties on $\alps(Q)$ 
from renormalisation scale variations \cite{Aktas:2005tz}.}
\label{f16}
\end{figure}

In case of the H1 analysis, the assumption of universal power corrections is assumed to be valid,
and a fit as a function of $Q$, according to the renormalisation group equation is performed,
 with the resulting scale dependence of $\alp(Q)$ shown in Fig.~\ref{f16}.

\subsection{Jet Rates}
\begin{figure}
\includegraphics[width=64mm]{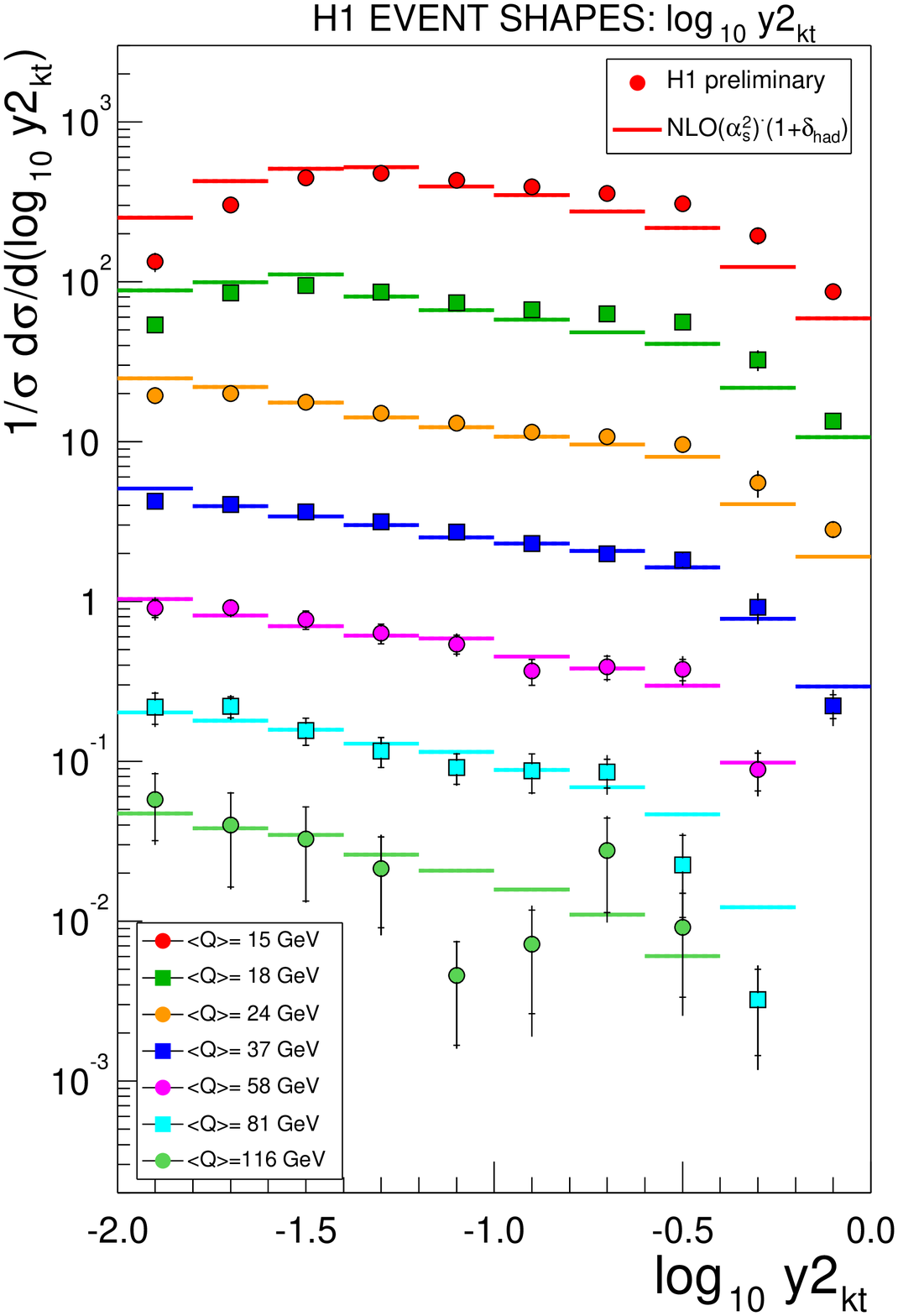}
\includegraphics[width=64mm]{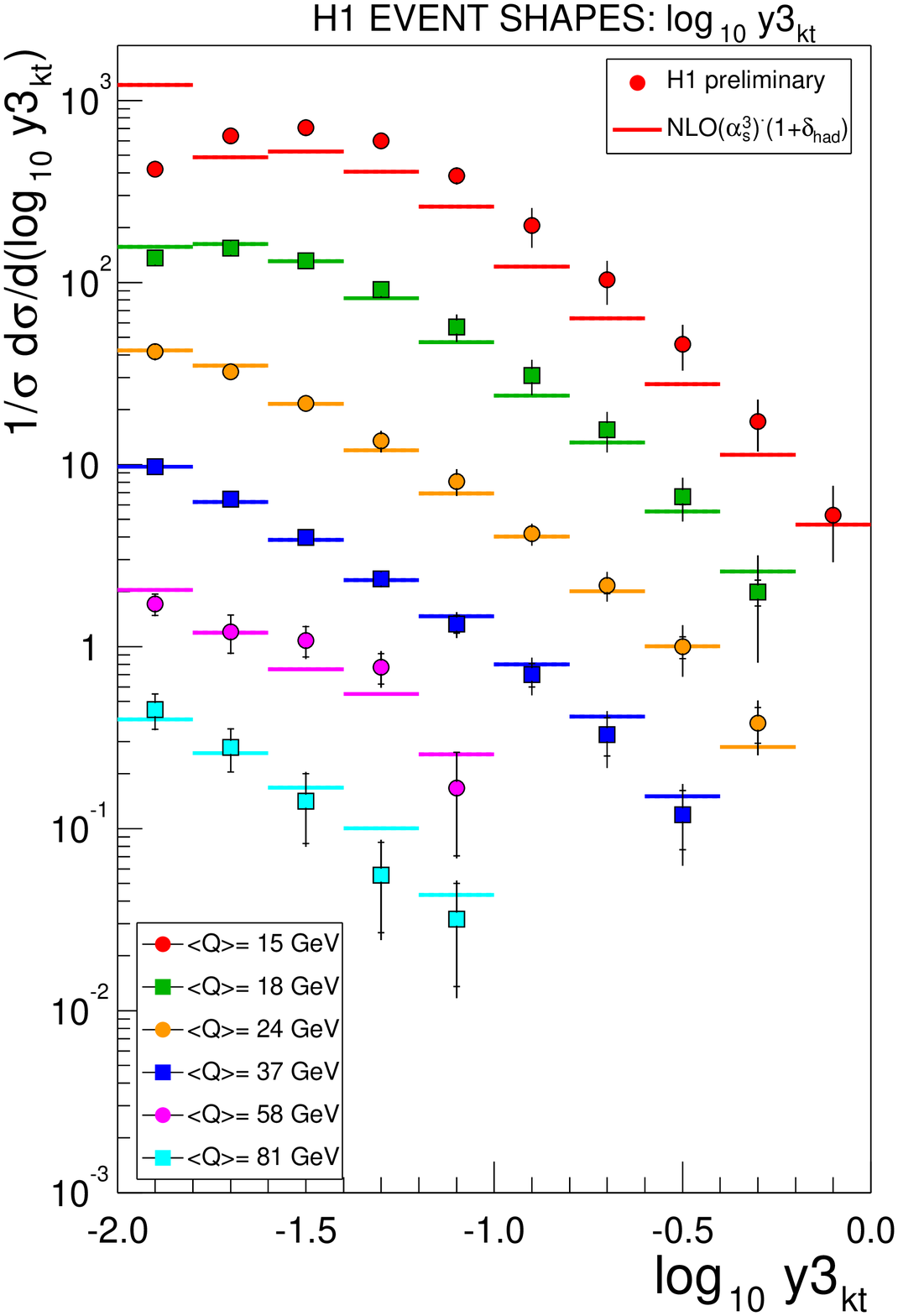}
\caption{The jet rate $y_2$ (left) and $y_3$ (right) \cite{Kluge:2003sa} compared to a calculation based on NLOJET++, where hadronisation corrections are determined with RAPGAP.}
\label{f13}
\end{figure}
The longitudinally invariant $k_t$ algorithm is the de-facto standard for 
jet analyses in DIS.
For this algorithm in its exclusive mode, the H1 Collaboration determined distributions of 2-jet and 3-jet rates  \cite{Kluge:2003sa}
as a function of $Q$, shown in Fig.~\ref{f13}.
Since the power corrections to this observables are not yet calculated, the data are compared to a NLO calculation,
corrected for hadronisation effects by the Lund string model as implemented in the RAPGAP \cite{Jung:1995gf} event generator.
At scales $Q>25\GeV$ a good description is provided for both jet rates, only at lower values deviations are
found, which might be due to insufficient treatment of the hadronisation.
However, as for the 2-jet event shapes, a soft gluon resummation matched to the fixed order part, might be important.
Unfortunately, such a calculation is not available yet.

\subsection{3-jet Event Shapes}
\begin{figure}
\includegraphics[width=90mm]{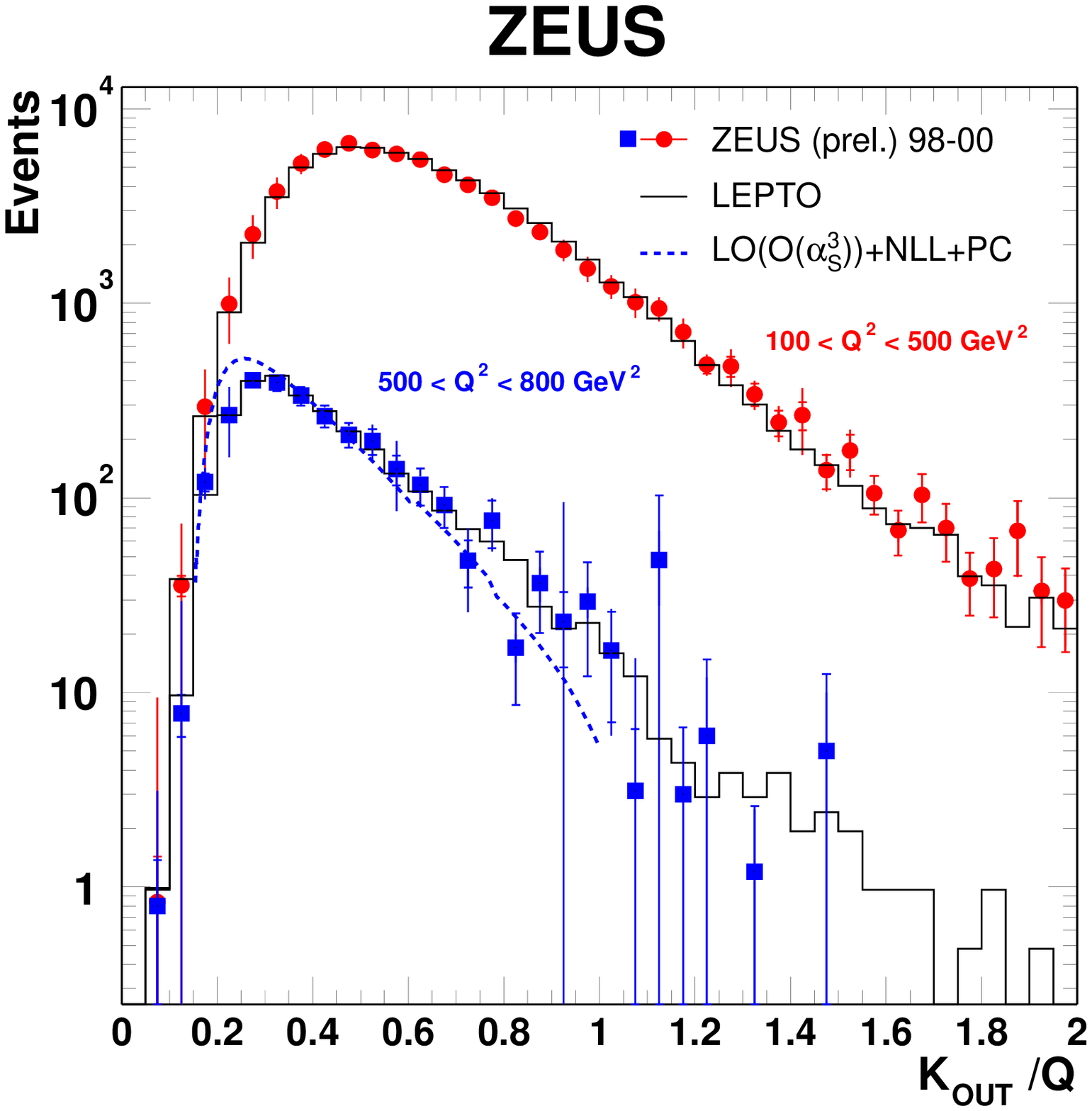}
\caption{$K_{out}$ distribution compared to LEPTO and the LO+NLL+PC calculation shown in bins of \mbox{$100\GeV^2<Q^2<500\GeV^2$}
and {$500\GeV^2<Q^2<800\GeV^2$} \cite{Everett:2004fg}.}
\label{f14}
\end{figure}

\begin{figure}
\includegraphics[width=80mm]{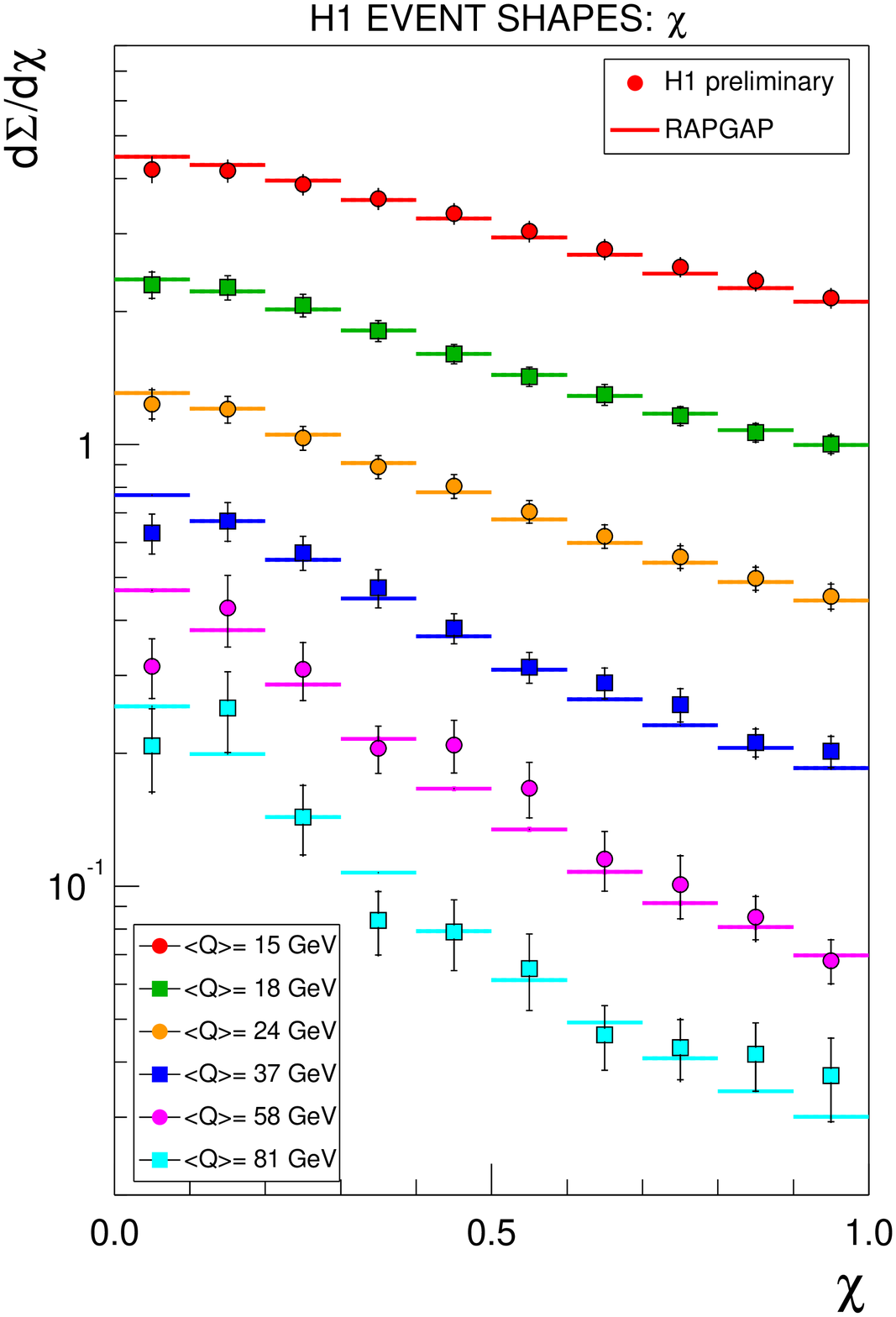}
\caption{Measured values of the azimuthal correlation  \cite{Kluge:2003sa}. The data are compared with results from the RAPGAP event generator.}
\label{f15}
\end{figure}

Compared to 2-jet event shape, the 3-jet variables differ in that they are
sensitive to large angle emissions between hard jets.
Also it has been proposed to use such variables in a hadron-hadron collision environment.
Preliminary measurements from the H1 \cite{Kluge:2003sa} and ZEUS \cite{Everett:2004fg} Collaborations are shown in Fig.~\ref{f14} and \ref{f15}.
The theory predictions are not yet on the level of precision as for the 2-jet variables.
A rather good description is obtained for the azimuthal correlation by the RAPGAP event generator,
 which relies on parton showers and the Lund string model. 
The out-of-plane momentum is compared to a LO+NLL calculation with power corrections,
 which does not a too bad job at high $Q^2$.
Some shift with respect to the data is seen, but one should keep in mind that the fixed order
part of the calculation is only at leading order.
As DIS data of good precision is available for both observables, 
the calculation at NLO+NLL precision would be highly desirable in order
to draw further conclusions about power corrections.

\section{Summary}
Concerning the mean values of 2-jet event shapes overall consistent findings
are reported by H1 and ZEUS, with support for universal
power corrections.
A rather large spread in the fitted values of  $\alp(m_Z)$ 
and the large scale uncertainties of the fitted results
suggest that the currently available fixed order calculations at NLO 
are not sufficient for this application.

For distributions of 2-jet event shapes soft gluon resummations 
are being used.
In a recent H1 publication the fits to the spectra resulted in 
compatible values for $\alpha_0$ and the strong coupling among the
five studied observables.
In addition the obtained values of $\alp(m_Z)$ were found to be
consistent with the world mean. 
A somewhat larger spread in  $\alpha_0$ is reported by a ZEUS preliminary result,
 but again the fitted  $\alp(m_Z)$ are compatible with the world mean.
Discrepancies in the findings of the two HERA experiments can be at least partly
attributed to uncertainties due to determination of the fit intervals and
due to differing matching prescriptions (between the fixed order the resummed calculations),
as the measured data appear to be compatible with each other.
Within the H1 analysis it was shown that the results were not sensitive to changes in the
fit interval.

The topic of distributions from 3-jet event shapes is 
experimentally more difficult, because of reduced statistics and 
 declined detector resolution.
However, as good DIS data are on the market for two of these variables, 
a perturbative NLO+NLL could, when available, allow for a stringent
test of power corrections beyond 2-jet variables.
In case of the jet rates, besides a NLO+NLL calculation more insight into 
the power correction coefficients is eligible to draw further conclusions. 
  
On the other hand, the observed universality of $\alpha_0$ for both
 distributions and mean values of the 2-jet event shape variables
 gives already strong support for the concept of power corrections.

NB: After this workshop the ZEUS Collaboration made available new results on event shapes \cite{Chekanov:2006hv}.
The main characteristics of the fitted results have not changed considerably compared to the 
preliminary results shown in this article. 

\subsection*{Acknowledgments} \label{Ack}
I would like to thank Mrinal Dasgupta, Yuri Dokshitzer and Gavin Salam
for the organisation of this stimulating workshop.  
Also I would like to thank Hans-Ulrich Martyn for useful comments on this manuscript
and for the collaboration on the subject of event shapes over the last years.

\bibliography{tkbib.bib}

\end{document}